\pdfoutput=1
\documentclass[aps,prl,floatfix,twocolumn,reprint,superscriptaddress]{revtex4}
\usepackage{amsfonts}
\usepackage{mathrsfs}
\usepackage{amsmath}
\usepackage{color}
\usepackage{graphicx}
\usepackage{bm}
\usepackage{amssymb}
\usepackage{xspace}
\usepackage{epstopdf}
\usepackage{dcolumn}
\usepackage{longtable}
\usepackage[colorlinks=true, letterpaper=true, pdfstartview=FitV, linkcolor=blue, citecolor=blue, urlcolor=blue]{hyperref}

\begin{document}
\title{ van der Waals Heterostructures of Germanene, Stanene and Silicene with Hexagonal Boron Nitride and Their Topological Domain Walls}

\author{Maoyuan Wang}
\affiliation{School of Physics, Beijing Institute of Technology, Beijing 100081, China}

\author{Liping Liu}
\affiliation{School of Physics, Beijing Institute of Technology, Beijing 100081, China}

\author{Cheng-Cheng Liu}
\email{ccliu@bit.edu.cn}
\affiliation{School of Physics, Beijing Institute of Technology, Beijing 100081, China}

\author{Yugui Yao}
\email{ygyao@bit.edu.cn}
\affiliation{School of Physics, Beijing Institute of Technology, Beijing 100081, China}

\begin{abstract}
We investigate van der Waals (vdW) heterostructures made of germanene, stanene or silicene with hexagonal Boron Nitride (h-BN). The intriguing topological properties of these buckled honeycomb materials can be maintained and further engineered in the heterostructures, where the competition between the substrate effect and external electric fields can be used to control the tunable topological phase transitions. Using such heterostructures as building blocks, various vdW topological domain walls (DW) are designed, along which there exist valley polarized quantum spin Hall edge states or valley-contrasting edge states which are protected by valley(spin)- resolved topological charges and can be tailored by the patterning of the heterojunctions and by external fields.

\end{abstract}
\maketitle

\section{INTRODUCTION}
Two-dimensional (2D) honeycomb layered materials and their designer van der Waals (vdW) heterostructures have attracted tremendous interest in material science and condensed matter physics since the mechanical exfoliation of graphene from graphite~\cite{Geim2013}.
In such kind of honeycomb materials, besides real spin, valley ($K$ or $K^\prime$) and sublattice (A or B) provide other tunable binary degrees of freedom to engineer their electronic properties, such as the remarkable quantum X (X=anomalous, spin and valley) Hall effects~\cite{Haldane1988,Kane2005a,Kane2005b,Xiao2007,Martin2008,yao2009,zhang2011,zhang2013,Ezawa2013prb,Pan2014,Pan2015prb,Gorbachev2014}.
Among these novel 2D honeycomb materials, silicene, germanene, and stanene with stable buckled honeycomb structures~\cite{Guzman-Verri2007,Cahangirov2009,Liu2} are attractive, and are predicted as the simplest elemental quantum spin Hall (QSH) insulators with sizable bulk gaps especially ~\cite{Liu1,Liu2}.
So far, lots of theoretical efforts have been devoted to engineer their novel electronic properties by various means, eg. with confinement~\cite{Ding2009a}, electric field~\cite{Ezawa2012a,Ni2012,Drummond2012,Tsai2013}, magnetic field~\cite{Zhang2013}, exchange field~\cite{Ezawa2012a,Ezawa2012b,Pan2014,Pan2015prb,Pan2015}, light fields~\cite{Ezawa2013}, domain walls~\cite{Ezawa2013prb,Pan2015} and chemical modifications~\cite{Xu2013}. In addition, there are several theoretical studies on the substrate effects for these 2D materials, such as on graphene~\cite{Cai2013}, hexagonal boron nitride (h-BN)~\cite{Kalonii2013,Liu2013,Li2013}, and between bilayer graphene~\cite{Neek-Amal2013} etc., whereas the impacts of the substrates on the topological properties are seldomly addressed.

On the experimental side, the three honeycomb materials have been synthesized on various substrates. For example silicene has been epitaxially grown on Ag(111)~\cite{Tao2015,Vogt2012,Feng2012,Chen2012,Lin2013}, Zr$_2$B$_2$(0001)~\cite{Fleurence2012}, Ir(111)~\cite{Meng2013}, Au(110)~\cite{Tchalala2013} and MoS$_2$ substrates~\cite{Chiappe2014}. Germanene has been manufactured on Pt(111)~\cite{Li2014}, Au(111)~\cite{Davila2014} and Al(111)~\cite{Derivaz2015} substrates. Stanene have recently been prepared on Bi$_2$Te$_3$(0001)~\cite{Zhu2015} and in liquid ambience~\cite{Saxena2015}. However on these substrates, the 2D honeycomb layers are prone to form various crystallographic reconstructions, with structures different from their pristine forms.
On the other hand, their distinct electronic properties around the Fermi level and their topological properties may be destroyed due to the hybridization with the substrates~\cite{Guo2013}.
Hence, it is crucial to find high-quality substrates that can stabilize these 2D materials and protest their novel topological properties, further to enable the engineering of their  topological properties in simple and feasible manners.


In this paper, based on density functional theory (DFT) and model analysis, we investigate two kinds of heterostructures of germanene, silicene and stanene monolayers and h-BN as well as domain walls (DWs) between different heterostructures. We find a new universal mechanism that in these buckled materials the competition and synergy between the substrate effect, heterostructures pattern and electric fields can be used to tune the topological phase transitions. Our results show that the h-BN substrate has little hybridization with the host materials. For symmetric heterostructures, the h-BN encapsulation protect the atomic structures and the electronic properties of germanene and stanenep, particularly maintaining their nontrivial bandgaps. For asymmetric structures, the effect of the h-BN substrate is to provide an effective staggered potential which could be compensated by an external electric field. Furthermore, we propose realizations of novel valley polarized QSH edge states based on a helical DW between an asymmetric heterostructure and a symmetric one, and valley-contrasting edge states along a DW between two inverse asymmetric heterostructures. The topological properties, such as valley and spin index, as well as the directionality and the numbers of the edge states are protected by topological charges and can be tailored by the patterning of heterojuctions and the external fields.  Devices with such tunable disspationless topological edge states have potential applications in spintroincs and valleytronics.

The DFT calculations are performed using the projector augmented wave method implemented in VASP~\cite{Kresse1996}. Perdew-Burke-Ernzerhof parametrization of the generalized gradient approximation (GGA-PBE) is used for the exchange correlation potential~\cite{Perdew1996,KJ99}. The energy cutoff of the plane wave basis is set to 500 eV and a $5\times5\times1$ kpoints mesh is used. The supercell structure was optimized until the force on each atom was less than 0.01eV/\AA\, and with the vdW correction (DFT-D2)~\cite{Klimes2011}. We then use an iterative method~\cite{Sancho} to obtain the surface Green's functions of DW systems, from which we calculate the dispersions of the topological edge states.

\subsection{\bf   Asymmetric and symmetric heterostructures}

\begin{figure}
\includegraphics[width=1.0\columnwidth]{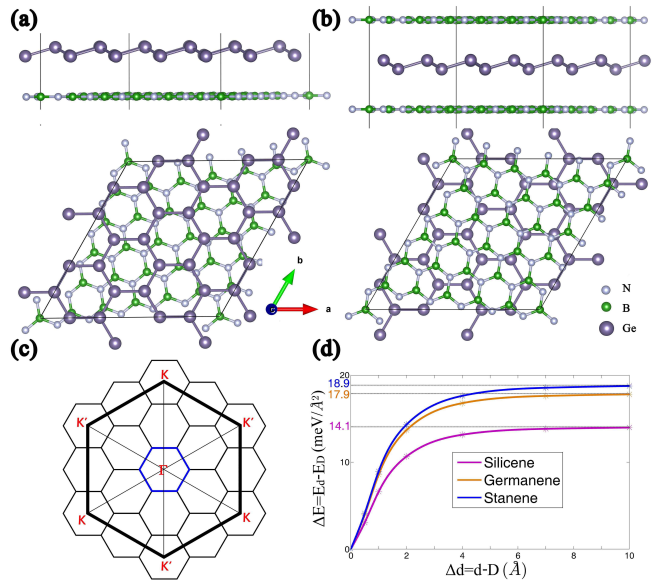}
\caption{{\bf Crystal structure of asymmetric and symmetric heterostructures.}
(a) The crystal structure of asymmetric heterostructures. The upper and lower panels are side view and top view. (b) Similar as (a) but for symmetric ones. (c) The Brillouin zone of the given supercell (small blue one) and primitive cell of germanene (large black one). $K$ and $K^\prime$ point of Brillouin zone are folded onto $\Gamma$ point. (d) The binding energy curves for silicene, germanene, and stanene on the h-BN substrate, where $E_D$ and $E_d$ are the energy of the asymmetric heterostructures  with the optimized interlayer distance D and the stretched one d. }

\label{fig1}
\end{figure}

Two types of heterostructures consisting of the host materials (silicene, germanene, stanene) and the h-BN substrate are taken into account. One is asymmetric heterostructures made of the hosts on the h-BN substrate (Fig.~\ref{fig1}(a)). Their geometry and corresponding interlayer distances are determined by employing the atomic structure optimization with the vdW corrections (DFT-D2). Their interlayer distances and binding energies are given in Table~\ref{TBparameter}, which are comparable to the interlayer distances and binding energies of the typical vdW materials; for instance 12 meV/\AA$^2$ in graphite and 26 meV/\AA$^2$  in MoS$_2$~\cite{Liu2012,Zhou2014}. Therefore these heterostructures are vdW-type structures. Our band structure calculations show the bands of h-BN substrate are far from the Fermi level, hence the h-BN substrate does not hybrid with the relevant low-energy levels of the hosts, but opens a trivial gap at the Dirac point without SOC. Moreover, according to our calculation, with more layers of h-BN under these host materials, the band gap will not change too much ($\sim$1meV), so we use a monolayer of h-BN as the substrate and the results could apply to the case of the several layers substrate.

\begin{table}
  \caption{Tight-binding (TB) parameters ($t$, $t_{so}$, $U_{h-BN}$) by fitting DFT and binding energies ($E_{b}$), interlayer distances ($D$) of the hosts and h-BN as well as the topological invariants ($Z_2$, $C_v$) for the three asymmetric heterostructures. As for TB parameters for symmetric heterostructures, $U_{h-BN}$ could be taken as 0 and others ($t$, $t_{so}$) could be the same as the asymmetric ones. }
  \label{TBparameter}
  \begin{tabular}{cccccccc}
    \hline
    \hline
    2D  & $t$   &$t_{so}$ & $U_{h-BN}$ & $E_{b}$   &   $D$ &   $Z_2$ &   $C_v$ \\
    Host material   &   (eV)   & (meV)  &    (meV)     &   (meV/\AA$^2$) &  (\AA)  &  & \\  \hline
    $Silicene$~\cite{Silicene}   & 1.07   & 0.78  & -16.1  & 14.1 &   3.36  &  0  & 2  \\
    $Germanene$ & 1.03  & 11.9  & -22.3 & 17.9 &   3.23 &  0  & 2 \\
    $Stanene$~\cite{Stanene}  & 0.93  &36.8   & -35.2 & 18.9 &   3.28  &  1  & 0\\
    \hline
  \end{tabular}
\end{table}

In the following, we take germanene as an example. As shown in Fig.~\ref{fig1}(a), the supercell, $2\sqrt{3}\times2\sqrt{3}$ germanene on $\sqrt{31}\times\sqrt{31}$  h-BN with a rotation angle 21$^{\circ}$, is chosen with 24 Ge atoms and with approximately 0.6 \% lattice mismatch.~\cite{Li2013} The two original  Dirac points $K$ and $K^\prime$ in Brillioun zone are folded onto the $\Gamma$ point for $2\sqrt{3}\times2\sqrt{3}$ structure as plotted in Fig.~\ref{fig1}(c)~\cite{5x5}.

\begin{figure*}
\includegraphics[width=2.0\columnwidth]{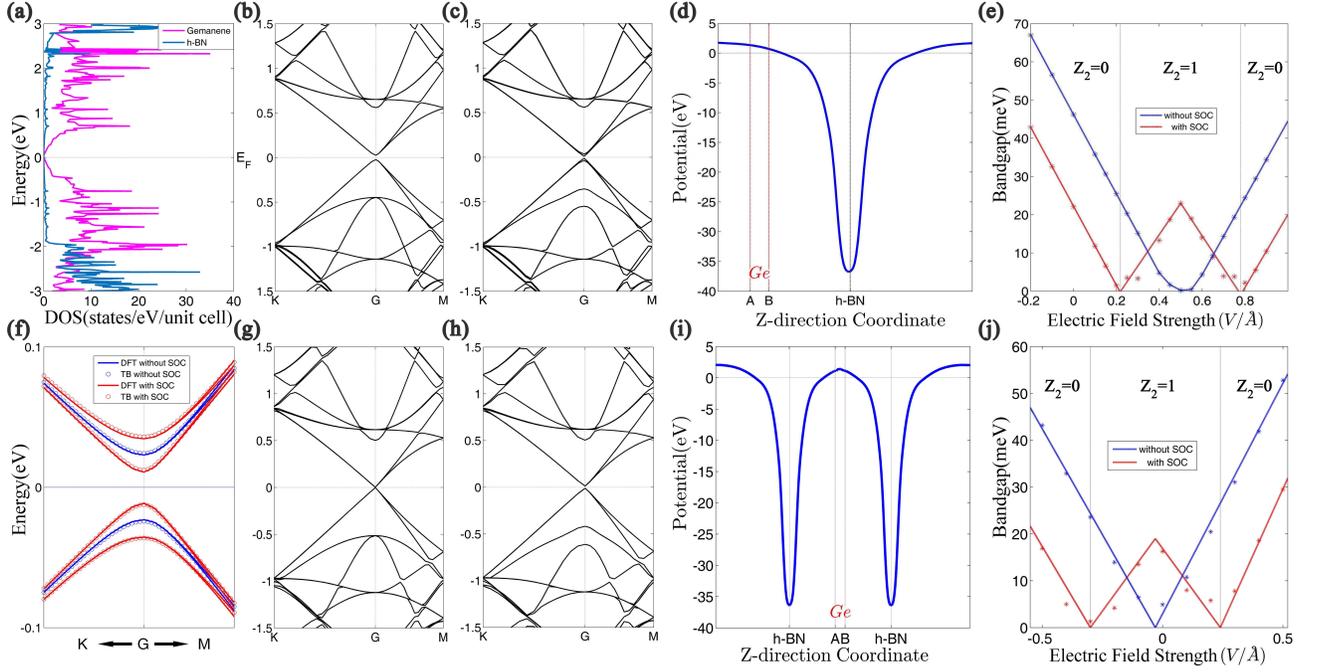}
\caption{{\bf Electronic structure of Asymmetric and symmetric heterostructures.}
(a)-(f) for asymmetric heterostructures with germanene on h-BN. (a) Density of states (DOS) without SOC. (b)(c) Band structures without and with SOC. They corresponding band gaps are 44.6 and 22.1 meV. (d) Potential energy distribution of the freestanding h-BN in real space. A and B represent the positions of Germanene sublattices, potentials of which are different. (e) Phase diagram in the plane of an external electric field and band gap. The red (blue) dots are the results achieved by DFT in the presence (absence) of SOC. Lines are provided to guide the eye. The topological number $Z_2$ is used to characterized the topological different phases for time-reversal invariant electronic systems ,and $Z_2=1$ means topological nontrivial whereas $Z_2=0$ indicates topological trivial. (f) The TB fitting with DFT for the SOC case (red lines and dots)  and without SOC case (blue lines and dots). (g)-(j) for symmetric heterostructures with germanene intercalating bilayer h-BN. (g)(h) are same as (b)-(c) but for the symmetric ones. (i) Same as (d) but for the symmetric heterostructures. There are almost the same potential at A and B sublattices. (j) Same as (e) but for the symmetric ones.}
\label{fig2}
\end{figure*}

Fig.~\ref{fig2} (a)(b) plot the density of states and band structure of the two-layer system without SOC. One can see that the bands around Fermi level come from germanene (red)(Fig.~\ref{fig2}(a)), and the bands of h-BN (blue) are far from Fermi level. As aforementioned, the two original Dirac points $K$ and $K^\prime$ with a gap about 44.6 meV are folded onto $\Gamma$ point. And when we remove h-BN out of the system, the profile of band structures will not change except for the gap closed in the freestanding germanene demonstrating that the role of the substrate will exerts the hosts a staggered potential. To understand the origin of the staggered potential, we calculate the potential energy distribution along $\bf{z}$ direction (Fig.~\ref{fig2}(d)). In the buckled germanene, the A and B sublattices feel different potential, responsible for a staggered potential. Because of breaking the inversion symmetry, the double degeneracy of the SOC band gap is lifted (Fig.~\ref{fig2}(c)). From Fig.~\ref{fig2}(e), the staggered potential can be counteracted by applying a perpendicular electric field of about 0.45 V/\AA, that means the h-BN substrate is equivalent to providing a perpendicular external electric field of -0.45 V/\AA. Without SOC, the band gap varies with a perpendicular external electric field in a V shape, and centered at 0.45 V/\AA. For the SOC case, we give a phase diagram, where topological phase transitions occur twice. The band gap varies with the electric field in a W shape, and also centered at 0.45 V/\AA. Around the center of the W are QSH insulators, whereas far from the center are band insulators.

The other is symmetric heterostructures, which has a sandwich structure by intercalating these hosts into the bilayer h-BN, plotted in Fig.~\ref{fig1}(b). Since the A and B sublattices of germanene feel almost the same potential energy in such symmetric sandwich structure (Fig.~\ref{fig2}(i)) unlike the asymmetric one, it is expected that its band strucutre is almost gapless when SOC is not taken into account (Fig.~\ref{fig2}(g)), and has a similar sizable SOC gap for SOC case (Fig.~\ref{fig2}(h)) as that of the pristine germanene. In the phase diagram (Fig.~\ref{fig2}(j)), it also has the similar W shape or V shape electric field response curves with or without SOC as the above asymmetric heterostructures case, but with the centers shifted slightly from the zero electric field. These features are almost the same as the pristine germanene. The advantages of the sandwich structure are that it is not only stabilize the atomic geometry of the host materials but also protect the intriguing electronic properties from the undesired influence of the substrates.

The physics of the two types of  asymmetric and symmetric heterostructures can be captured by a TB lattice model~\cite{Liu2,Ezawa2012a},
\begin{equation}\label{TB}
\begin{split}
H=& -t\sum_{\left\langle ij\right\rangle \alpha}c_{i\alpha}^{\dagger}c_{j\alpha}+\frac{it_{SO}}{3\sqrt{3}}\sum_{\left\langle \left\langle ij\right\rangle \right\rangle \alpha\beta}\nu_{ij}c_{i\alpha}^{\dagger}\sigma_{\alpha\beta}^{z}c_{j\beta} \\
& +U_{h-BN}\sum_{i\alpha}c_{i\alpha}^{\dagger}\sigma_{ii}^{z}c_{j\alpha}+U_{EF}\sum_{i\alpha}c_{i\alpha}^{\dagger}\sigma_{ii}^{z}c_{j\alpha}.
\end{split}
\end{equation}
The first term is the nearest hopping term, and the second term is the intrinsic first-order SOC term. The last two terms are the staggered potential due to h-BN substrates and an external electric field. These parameters for the asymmetric heterostructures are fitted with DFT band structures, as given in Table~\ref{TBparameter}. As plotted in Fig.~\ref{fig2}(f) for germanene, the TB fitting with DFT agrees with each other very well for both cases of SOC or not. And for the symmetric heterostructures, it is a good approximation taking $U_{h-BN}$ as zero.

We are interested in the low-energy physics and consider the effects of the h-BN on the host materials effectively providing a staggered potential for pristine them. Hence we expand the TB Hamiltonian surrounding the two valley $K$ and $K^\prime$, and obtain the low-energy effective model
\begin{equation}\label{Heff} H_{eff}=v_{F}\left(\tau_{z}k_{x}\sigma_{x}-k_{y}\sigma_{y}\right)+\left(M_{\tau_{z},s_{z}}+U_{EF}\right)\sigma_{z},
\end{equation}
with mass term $M_{\tau_{z},s_{z}}\equiv t_{SO}\tau_{z}s_{z}+U_{h-BN}$, Fermi velocity $v_{F}=\sqrt{3}/2at$, and Pauli matrix $\bf{\tau}$, $\bf{s}$, and $\bf{\sigma}$ acting in the space of valley, spin, and sublattice space, respectively. Since valley and spin are conserved here, for an explicit valley and spin, the corresponding projected topological charges are given~\cite{formula},
\begin{equation}\label{TC}
C\left(\tau_{z},s_{z}\right)=-\frac{\tau_{z}}{2}sgn\left(M_{\tau_{z},s_{z}}+U_{EF}\right).
\end{equation}
From Eq.~\ref{TC}, we can readily obtain the total Chern numbers, spin Chern numbers, $Z_2$, and valley Chern numbers written as $C=\sum_{\tau_{z},s_{z}}C\left(\tau_{z},s_{z}\right)$, $C_{\uparrow/\downarrow}=\sum_{\tau_{z}}C\left(\tau_{z},s_{z}=\uparrow/\downarrow\right)$, $Z_{2}=\left(C_{\uparrow}-C_{\downarrow}\right)/2$ mod 2, $C_{K/K'}=\sum_{s_{z}}C\left(\tau_{z}=K/K',s_{z}\right)$, $C_{v}=C_{K}-C_{K'}$. Based on these formulae,  for the above asymmetric heterostructures, the corresponding Chern numbers are given in Table~\ref{TBparameter}.

\subsection{\bf   Topological domain walls}
\begin{figure}
\includegraphics[width=1.0\columnwidth]{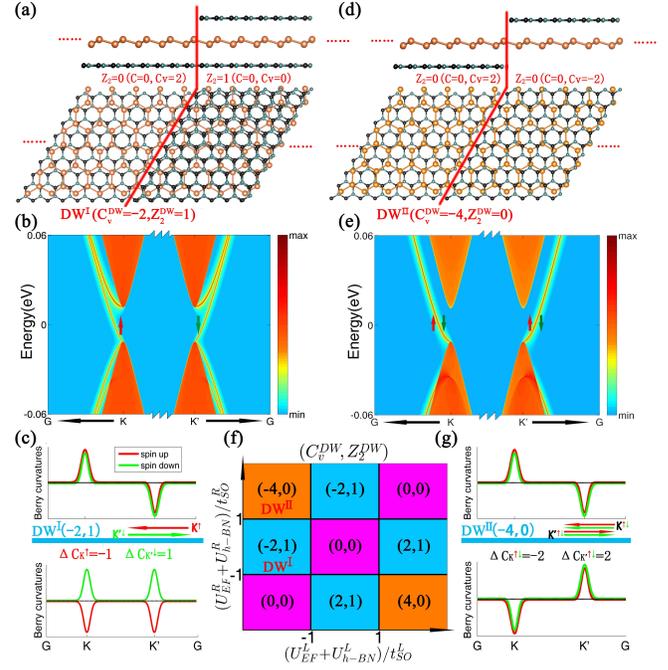}
\caption{{\bf Topogical domain walls}
(a) DW1 system made of the semi-infinite asymmetric and symmetric heterostructures. The upper and lower panels are side view and top view. (b) The energy spectrum for the DW exhibiting valley-polarized QSH edges. (c) The schematic diagram for the Berry curvature distribution for the both sides of the DW in the respective upper and lower panels, and the valley-polarized QSH edges along the DW in the middle panel. (d)(e)(g) The similar as (a)-(c) but for DW2 system made of the two inverse semi-infinite asymmetric heterostructures. Along the DW are valley-contrasted edges states. (f) The phase diagram for a general DW system.}
\label{fig3}
\end{figure}

We propose two novel topological DW systems that can be considered as composite horizonal heterostructures, as illustrated in Fig.~\ref{fig3}(a)(d). One consists of the above asymmetric and symmetric heterostructures, and the other consists of the two inverse asymmetric heterostructures.
By exploiting the surface Green method~\cite{Sancho1985}, we plot the spectra for the both zigzag DW systems. Notice that the following analysis from the topological view  is not limited to the zigzag case but also applies to all the DW systems with two differentiable valleys along their edges. For the first DW (DW1), helical edge states emerge in the bulk gap with one (the other) edge encoded spin up (down) around $K$ ($K^\prime$) moving left (right)(Fig.~\ref{fig3}(b)), which is reminiscent of the edge states of QSH effects, but valley-polarized here. The emerging of such novel helical DW edge states can be understood in terms of the above topological charges formulae (Eq.~\ref{TC}). The topological charges for the DW systems read
\begin{equation}\label{DW_TC}
\begin{split}
& C^{DW}\left(\tau_{z},s_{z}\right) =C^{R}\left(\tau_{z},s_{z}\right)-C^{L}\left(\tau_{z},s_{z}\right)\\
&=\frac{\tau_{z}}{2}\left[sgn\left(M_{\tau_{z},s_{z}}^{L}+U_{EF}^{L}\right)-sgn\left(M_{\tau_{z},s_{z}}^{R}+U_{EF}^{R}\right)\right],
\end{split}
\end{equation}
where the superscripts $DW$, $L$, and $R$ stand for domain wall, and its left and right side. The Berry curvature distribution for the both sides are shown in the upper and lower panels of Fig~\ref{fig3}(c), respectively, and the corresponding Chern number are $C_{K\uparrow}^{L}=C_{K\downarrow}^{L}=1/2$, $C_{K'\uparrow}^{L}=C_{K'\downarrow}^{L}=-1/2$, $C_{K\uparrow}^{R}=-C_{K\downarrow}^{R}=-1/2$, and $C_{K'\uparrow}^{R}=-C_{K'\downarrow}^{R}=-1/2$. Therefore, the topological charges for the DW are $C_{K\uparrow}^{DW}=-1$ and $C_{K'\downarrow}^{DW}=1$, which agrees with the helical edge states in Fig~\ref{fig3}(b). Consequently, the topological invariants for DW1 are $C_{v}^{DW}=-2$ and $Z_{2}^{DW}=1$, as shown in the middle panel of Fig~\ref{fig3}(c), indicating valley-polarized QSH effects, which we name VP-QSH effects. This phase is also proposed in the Y-junctions by M. Ezawa~\cite{Ezawa2013prb}.

\begin{figure}
\includegraphics[width=1\columnwidth]{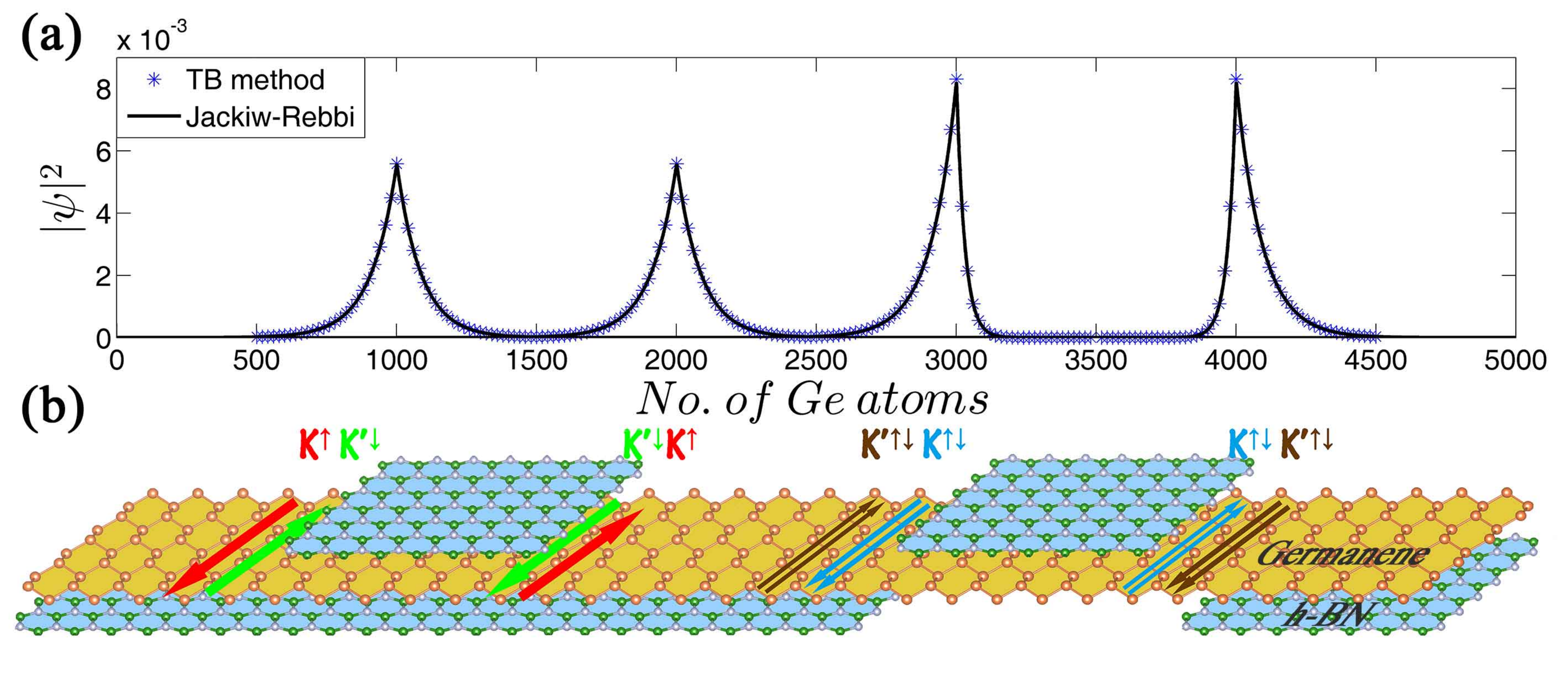}
\caption{{\bf Potential device applications}
Disspationless devices design of the DW edges of VP-QSH and high valley Chern number for parallel DWs. (a) The module square of the wavefunction distribution of the edge states for the parallel DWs in (b). The stars are from TB model, and the lines are the analytic Jackiw-Rebbi solutions.}
\label{fig4}
\end{figure}

For the second DW (DW2) (Fig~\ref{fig3}(d)), on its both side are the inverse asymmetric heterostructures. There are four valley-contrasted edge states in its bulk gap (Fig~\ref{fig3}(e)). Similar to the case of DW1, we can explain the valley-contrasted edge states with high valley Chern number by the direct calculation of the topological charges: $C_{K\uparrow}^{L}=C_{K\downarrow}^{L}=1/2$, $C_{K'\uparrow}^{L}=C_{K'\downarrow}^{L}=-1/2$, $C_{K\uparrow}^{R}=C_{K\downarrow}^{R}=-1/2$, $C_{K'\uparrow}^{R}=C_{K'\downarrow}^{R}=1/2$, and thus $C_{v}^{DW}=-4$ (Fig~\ref{fig3}(g)). Finally we given an entire phase diagram in the plane of $\left(U_{EF}^{L}+U_{h-BN}^{L}\right)/t_{SO}^{L}$-$\left(U_{EF}^{R}+U_{h-BN}^{R}\right)/t_{SO}^{R}$ with $L$ and $R$ labeling the left and right parts of the DW. As shown in Fig.~\ref{fig3}(f), there are four topological phases and one trivial phase, where DW1 and DW2 are placed at suitable positions. In addition, by applying respective gate voltages on the sides of a DW, topological phase transitions can occur.

The patterns in such kinds of DWs configurations are by virtue of engineering the h-BN substrates that could have a sharp edge cut~\cite{Alem2009,Gao2013}, whereas the topological edge states live in the host materials. Therefore, the DWs have clean topological edge states with much less dangling bonds than the nanoribbons of these hosts, and thus these topological edge states could be more beneficial for the application in spintroincs and valleytronics.

\section{\bf  Discussions}

The zero-energy edge states are exponentially localized near the DW, across which the sign of the mass term is inverted, i.e., $\left[M_{\tau_{z},s_{z}}\left(x<0\right)+U_{EF}\left(x<0\right)\right]\times \left[M_{\tau_{z},s_{z}}\left(x>0\right)+U_{EF}\left(x>0\right)\right]<0$. For the $K\uparrow$ channel of DW1 with dispersion $E=-v_{F}k_{y}$, its wavefunction is analytically given by the Jackiw-Rebbi solution~\cite{Jackiw1976},
\begin{equation}\label{J-B}
\psi_{k_y}\left(x,y\right)=\frac{1}{N}\left(\begin{array}{c}
1\\
i
\end{array}\right) \exp\left\{ ik_{y}y-\frac{1}{v_{F}}\int_{0}^{x}M\left(x'\right)dx'\right\},
\end{equation}
where $N$ is a normalization constant, $M\left(x<0\right)=U_{h-BN}^{L}+t_{SO}^{L}, M\left(x>0\right)=t_{SO}^{R}$, and $U_{EF}$ here is zero. For the other channels, the Jackiw-Rebbi solutions can also be determined in a similar way. We calculate these localized wavefunction distribution by using both analytical Jackiw-Rebbi solution (Eq.~\ref{J-B}) and TB model (Eq.~\ref{TB}), which agree with each other, as shown in Fig.~\ref{fig4}(a). Furthermore we propose a parallel DWs device concept based on the intrinsic two kinds of DWs, which could be tailored by different superlattice patterns. Meanwhile, a perpendicular external electric field is a good knob to control the intriguing DW edge states. In order to better utilize these DW edge states, and it is desired that the states located on different DWs do not overlap appreciably. For example, if we require the overlap to be as small as 1 \% of the state, then the distance of the DWs has a lower bound $L_{min}$, and $L_{min}/a=\frac{\sqrt{3}}{2}ln10\left(\frac{t^{L}}{\mid M^{L}\mid}+\frac{t^{R}}{\mid M^{R}\mid}\right)$ with $t$ the nearest neighbor hopping, M the mass terms. Take germanene+h-BN heterostructures in the absence of an external electric field as an example, $L_{min}^{DW1}\cong371 a\cong150 nm$ and $L_{min}^{DW2}\cong258 a\cong104 nm$~\cite{note3}.

For stanene and silicene plus h-BN heterostructures, we can perform the similar analysis as the case for germanene. Their TB parameters are obtained by fitting with DFT band structures, given in Table~\ref{TBparameter}. It can be found out that for the silicene heterostructures, the nontrivial gap of pristine silicene is much smaller than staggered potential in the asymmetric one and comparable with the slight staggered potential in the symmetric one, which implies that it could realize both the DW1 with the help of an external electronic field applied on the side of the symmetric one and the DW2 with the high Chern number edges states. As for the stanene case, the staggered potential of h-BN subtrate cannot make stanene own a trivial gap~\cite{Astanene}, hence both asymmetric and symmetric sandwich heterostructures are ideal systems to preserve the intriguing topological properties of stanene. Moreover,
for the asymmetric heterostructures, since the staggered potential of h-BN substrate is comparable to the SOC of stanene, an external perpendicular electric field that can achieve in recent experiment conditions can effectively tune the band gap and the topological properties. This might have potential applications in the topological field-effect transistors~\cite{Ezawa2013apl}.

In view of the success of epitaxial growth of single- domain graphene on h-BN~\cite{Yang2013}, it is promising to epitaxial grow germanene, stanene, and silicene on h-BN. Moreover, given a silicene field-effect transistor has been made experimentally by a smart fabrication process of growing silicene on Ag(111) and then transferring it to an insulating Al2O3/SiO2 substrate~\cite{Tao2015}, we believe that it is feasible to transfer the grown germanene, stanene, and silicene onto h-BN substrate.
Due to the slight lattice mismatch here, the 2D superlattices of Moir$\acute{e}$ pattern may emerge and are particularly appealing, e.g., the Hofstadter's butterfly and the fractal quantum Hall effect~\cite{Dean2013,Hunt2013}. Consequently all the systems we mention in this letter could be used as an excellent platform for fundamental scientific study and promising device application.

\section{acknowledgments}
The authors would like to thank Dr. Shengyuan A. Yang, Dr. Zhiming Yu and Dr. Jin-Jian Zhou for helpful discussions. This work was supported by the MOST Project of China (Nos. 2014CB920903, and 2013CB921903), the National Natural Science Foundation of China (Grant Nos. 11574029, 11404022, and 11225418), the Specialized Research Fund for the Doctoral Program of Higher Education of China (Grant No. 20121101110046), Excellent young scholars Research Fund (Grant No. 2014CX04028) and the Basic Research Funds (Grant No. 20141842001) of Beijing Institute of Technology, and International Graduate Exchange Program of Beijing Institute of Technology.

\section{appendix}

\appendix
\setcounter{figure}{0}
\renewcommand{\thefigure}{A\arabic{figure}}
\subsection{TB+Green Function Method}
To display band structures of domain walls, we use TB parameters got from FP calculation and Green Function Method~\cite{Sancho1985}:

\[\begin{array}{l}
G=\left( {\begin{array}{*{20}{c}}
{{G_L}}&{{G_{LD}}}&{{G_{LR}}}\\
{{G_{DL}}}&{{G_D}}&{{G_{DR}}}\\
{{G_{RL}}}&{{G_{RD}}}&{{G_R}}
\end{array}} \right)\\
 \\
= {\left( {\begin{array}{*{20}{c}}
{(\varepsilon  - {H_L})}&{{h_{LD}}}&{{h_{LR}}}\\
{{h_{DL}}}&{(\varepsilon  - {H_D})}&{{h_{DR}}}\\
{{h_{RL}}}&{{h_{RD}}}&{(\varepsilon  - {H_R})}
\end{array}} \right)^{ - 1}}\\
\\
{G_D} = {(\varepsilon  - {H_D} - {\Sigma _L} - {\Sigma _R})^{ - 1}}\\
\\
N(\varepsilon ) =  - \frac{1}{\pi }{\mathop{\rm Im}\nolimits} [Tr{G_D}(\varepsilon )]
\end{array}\]

In the first equation, the Green function ($G$) of total system is defined including left ($L$), right ($R$) and central domain wall ($D$) regions, which can be calculated from Hamiltonian ($H$, $h$).  And as for the Green function of central domain wall region, we can get it from the second equation after calculating the self-energies ($\Sigma$) from the left and right regions: ${\Sigma _{L(R)}} = {h_{DL(R)}}{g_{L(R)}}{h_{L(R)D}}$. Finally, the energy spectrum for domain wall can be calculated from the third equation.

\subsection{$5\times5$ germanene on h-BN}
\begin{figure}

\includegraphics[width=0.8\columnwidth]{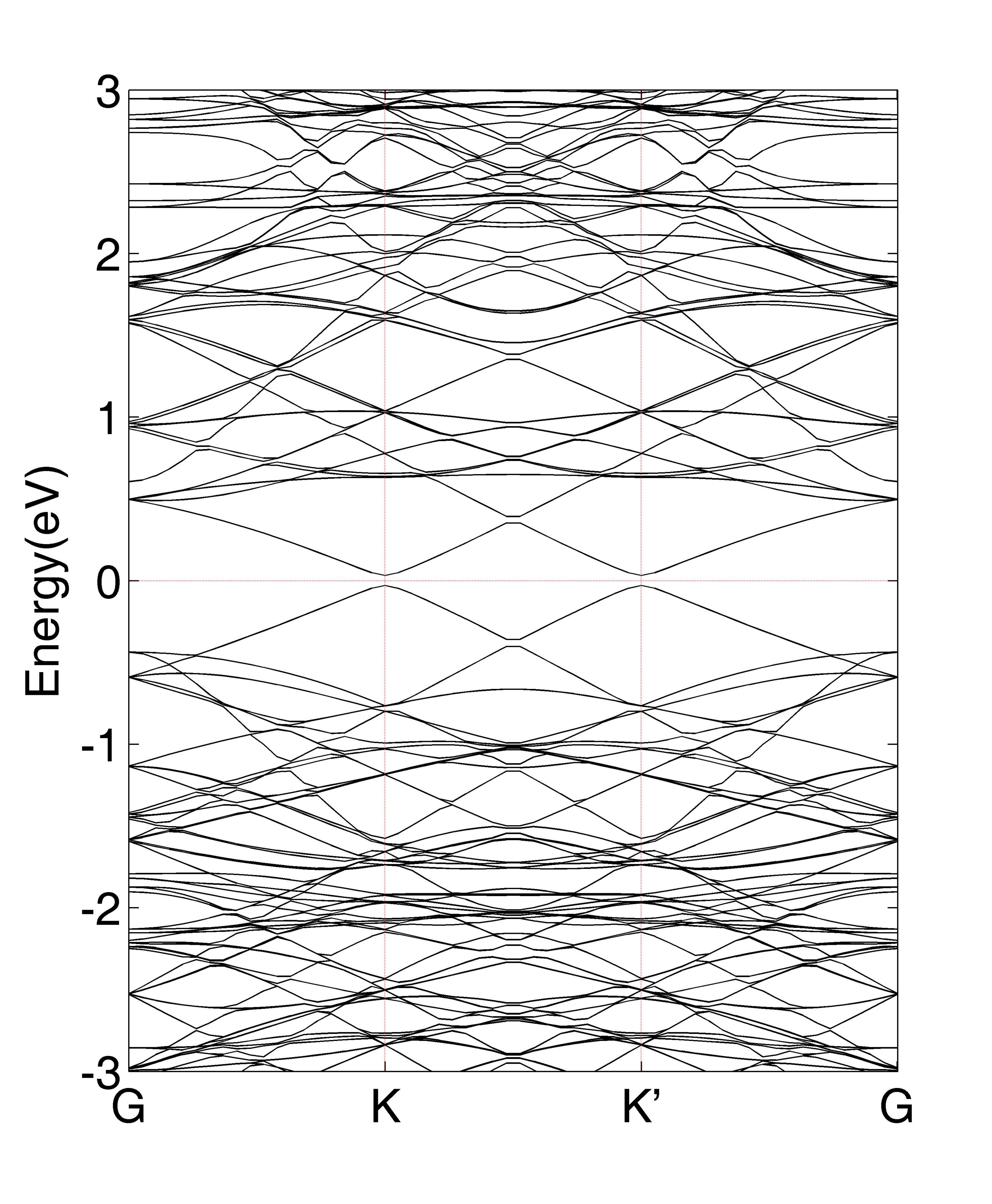}
\caption{\bf Bandstructure of asymmetric heterostructures with $5 \times 5$ germanene on h-BN without SOC.}
\label{figA1}
\end{figure}

In order to balance the accuracy and efficiency, we use the $2\sqrt{3}\times2\sqrt{3}$ supercell structure( $2\sqrt{3}\times2\sqrt{3}$ )germanene on $\sqrt{31}\times\sqrt{31}$ h-BN with a rotation angle 21$^{\circ}$, 24 Ge atoms and approximately 0.6\% lattice mismatch), where the valley $K$ and $K^\prime$ are folded onto the $\Gamma$ point. For the other case that the two Dirac points $K$ and $K^\prime$, which can be valley degree of freedom for manipulating information in valleytronics, are kept under the Brillioun zone folding. We also choose another larger supercell of 178 atoms (50 Ge), $5\times5$ germanene on $8\times 8$ h-BN with 0.15\% lattice mismatch. Its band structure with a gap of 61.6 meV is given in Fig.~\ref{figA1}. It should be noticed that even both the staggered potentials (-22.3 and -30.8 meV) is slightly different mainly due to the different buckling of different lattice mismatch, they give a reference interval of the staggered potentials which the h-BN substrates can provide. Moreover the reference interval is larger than the effective spin-orbit coupling strength, half the spin-orbit coupling gap, of Germanene (11.9 meV), hence will give the same topological properties.

\subsection{$3\times3$  stanene on h-BN}
\begin{figure}

\includegraphics[width=0.8\columnwidth]{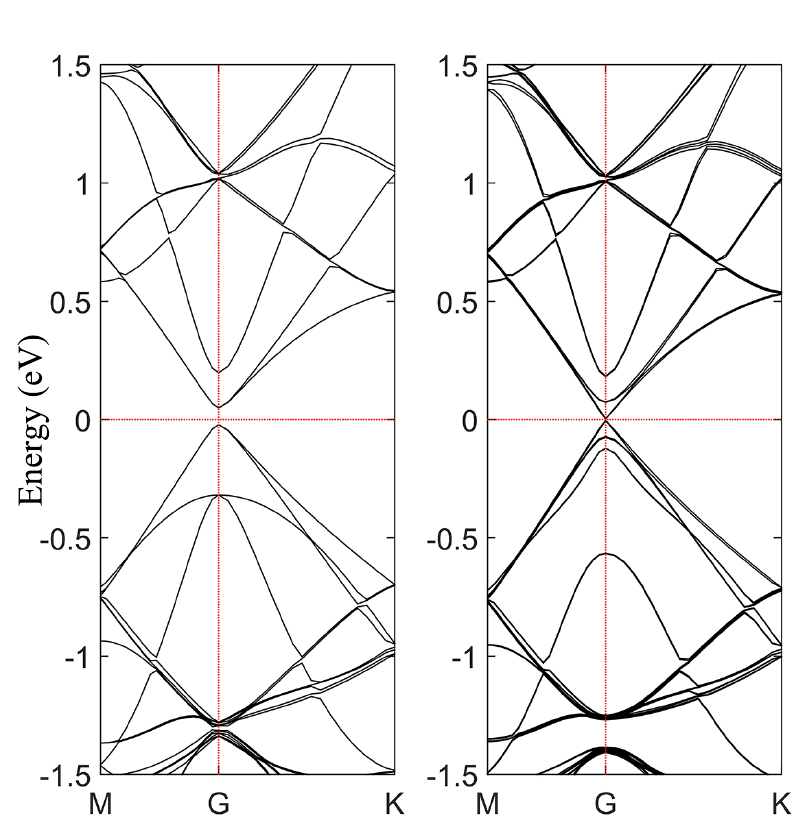}
\caption{\bf Bandstructure of asymmetric heterostructures with $3 \times 3$ stanene on h-BN without (a) and with (b) SOC.}
\label{figA2}
\end{figure}

As for the calculation of stanene, we choose $3\times3$ stanene on $\sqrt{31}\times\sqrt{31}$  h-BN with a rotation angle 9$^{\circ}$, with 18 Sn atoms and with approximately 0.25 \% lattice mismatch. The bandstructure of the asymmetric heterostructures without and with SOC are shown in Fig.~\ref{figA2}. The effective electric field of h-BN opens a trival gap about 70.4meV without SOC. When SOC taken into account, the asymmetric heterostructure of stanene is almost gapless with a small 3meV gap. From the TB parameters mentioned in Table~\ref{TBparameter}, it is obvious that $t_{so}>|U_{h-BN}|$, so the gap is a topological gap according to the TB model mentioned before.

\begin{thebibliography}{100}
\bibitem{Silicene} The supercell of DFT calculations for parameters here is $\sqrt{3}\times\sqrt{3}$ silicene on $\sqrt{7}\times\sqrt{7}$  h-BN with a rotation angle 10.9$^{\circ}$, which is chosen with 6 Si atoms and with approximately 0.67 \% lattice mismatch.
\bibitem{Stanene} As for the supercell of stanene, $3\times3$ stanene on $\sqrt{31}\times\sqrt{31}$  h-BN with a rotation angle 9$^{\circ}$, is chosen with 18 Sn atoms and with approximately 0.25 \% lattice mismatch.
\bibitem{note3}  To increase the integration level of the domain wall systems, it needs some conditions: a huge nontrival gap such as BiH and a huge trival gap caused by the intrinsic properties of materials or a large effective electric field induced by substrates.
\end{thebibliography}
\end{document}